\documentclass[pra,onecolumn,aps,superscriptaddress,notitlepage,longbibliography]{revtex4-1}
\PassOptionsToPackage{usenames,dvipsnames}{color}
\usepackage{amsmath,amsfonts,amssymb,amsthm,graphicx,bbm,enumerate,times,color,textcomp}
\usepackage{dsfont}
\usepackage{mathtools}
\usepackage{subcaption}

\newcommand{\mean}[1]{\langle #1 \rangle}
\newcommand{\cale}{\mathcal{E}}

\newcommand{\cals}{\mathcal{S}}
\newcommand{\calt}{\mathcal{T}}

\renewcommand\Re{\operatorname{Re}}

\usepackage{comment}

\definecolor{rodrigo}{rgb}{0,.4,1}

\definecolor{henrik}{rgb}{1,.4,0}

\newcommand{\mc}[1]{\mathcal{#1}}

\renewcommand{\Re}{\mathrm{Re}}

\newcommand{\tr}{\mathrm{Tr}} 
\newcommand{\Tr}{\mathrm{Tr}} 

\newcommand{\R}{\mb{R}}

\newcommand{\ket}[1]{\left.\left|{#1}\right.\right\rangle}

\newcommand{\bra}[1]{\left.\left\langle{#1}\right.\right|}

\newcommand{\ketbra}[2]{\ket{#1} \!\!\! \bra{#2}}
\newcommand{\proj}[1]{\ketbra{#1}{#1}}

\def\S{\sf S}
\def\B{\sf B}
\def\C{\sf C}
\def\R{\sf R}

\newcommand{\fu}{Dahlem Center for Complex Quantum Systems, Freie Universit{\"a}t Berlin, 14195 Berlin, Germany}

\usepackage{hyperref}
\begin{document}
\title{Cooling to absolute zero: The unattainability principle}
\author{Nahuel Freitas}
\affiliation{Theoretische Physik, Universit\"at des Saarlandes, D-66123 Saarbr\"ucken, Germany}
\author{Rodrigo Gallego}
\affiliation{\fu}
\author{Llu\'is Masanes}
\affiliation{Department of Physics \& Astronomy, University College London, WC1E 6BT London, U.K.}
\author{Juan Pablo Paz}
\affiliation{Departamento de F\'\i sica, FCEyN, UBA, Pabell\'on 1,
Ciudad Universitaria, 1428 Buenos Aires, Argentina}
\affiliation{Instituto de F\'\i sica de Buenos Aires, UBA CONICET,
Pabell\'on 1, Ciudad Universitaria, 1428 Buenos Aires, Argentina}

\begin{abstract}
The unattainability principle (UP) is an operational formulation of the third law of thermodynamics stating the impossibility to bring a system to its ground state in finite time. In this work, several recent derivations of the UP are presented, with a focus on the set of assumptions and allowed sets of operations under which the UP can be formally derived. First, we discuss derivations allowing for arbitrary unitary evolutions as the set of operations. There the aim is to provide fundamental bounds on the minimal achievable temperature, which are applicable with almost full generality. These bounds show that perfect cooling requires an infinite amount of a given resource---worst-case work, heat bath's size and dimensionality or non-equilibrium states among others---which can in turn be argued to imply that an infinite amount of time is required to access those resources. Secondly, we present derivations within a less general set of operations conceived to capture a broad class of currently available experimental settings. In particular, the UP is here derived within a model of linear and driven quantum refrigerators consisting on a network of harmonic oscillators coupled to several reservoirs at different temperatures.
\end{abstract}

\maketitle
\section{Introduction}\label{sec:intro}

The necessity of the third law of thermodynamics and its  physical content were heatedly debated by Nernst, Planck and Einstein at the beginning of the 20th century.
Several inequivalent formulations of the law~\cite{Nernst1906, Planck1911, Einstein1914, Nernst1906_2} were proposed, but the one that has been mostly considered by subsequent authors is the
\begin{quotation}\noindent
{\bf Unattainability principle:} \emph{It is impossible by any procedure, no matter how idealized, to reduce any assembly to absolute zero temperature in a finite number of operations} (Nernst~\cite{Nernst-1912}).
\end{quotation}
The above statement makes use of ambiguous concepts such as ``procedure'' and ``operation'' which are concomitant to formulations of thermodynamics present at the time.
Within the contemporary formulation of thermodynamics, by ``any procedure" it is meant any process whose underlying dynamics is unitary, and hence, it does not include measurements or preparations (unless the measurement apparatus is included in the ``assembly").
Note that, otherwise, we could violate the unattainability principle (UP) simply by measuring the energy of a two level system and conditionally driving it to its ground state.

Another ambiguity is the notion of ``operation", on which, supposedly, any procedure can be decomposed. A finite number of operations translates simply in that the duration of the overall procedure is finite.
This relation between finiteness of time and number of operations is reminiscent of the very specific type of thermodynamic operations (isothermal, adiabatic, etc.) considered at the time.
At present, however, we would like a formulation of the UP that applies to the widest range of physical procedures ---not necessarily decomposable into specific types of operations--- hence, we need to generalize the constraint that \emph{the time duration of the procedure is finite} in a setup-independent fashion.
This can be done with the following
\begin{quotation}\noindent
{\bf Finiteness assumption:} \emph{Within a finite time, a system can only interact with finitely-many other systems, each having effectively finite size. Also, within a finite time, only a finite amount of work can be injected into a system}.
\end{quotation}
The notion of ``finite size" that appears in the above assumption is formalized in different ways: finite heat capacity, finite volume (Sec.~\ref{sec:masanes}), finite Hilbert space dimension (Sec.~\ref{sec:single_inequality}) and finite largest eigenvalue of the Hamiltonian (Sec.~\ref{sec:dimension}).
The finiteness of the volume can be justified by, for example, invoking the finite speed of information propagation from Special Relativity or the Lieb-Robinson bound \cite{Lieb-Robinson}.
The finiteness of the Hilbert space dimension is more appropriate in the context of quantum computers and artificial systems.
This is particularly relevant because quantum computation requires initial pure states, and as we see below, the task of distilling pure states is essentially equivalent, in relation to the UP, to that of cooling to absolute zero.

The Finiteness Assumption also puts limits on the amount of thermodynamic resources (Sec.~\ref{sec:single_inequality}) and work (Secs.~\ref{sec:masanes} and \ref{sec:dimension}) that can be consumed in a cooling process.
The translation of ``time" to all these mentioned physical parameters allows to go beyond the original UP, and provide quantitative versions of it.
That is, relationships between the lowest achievable temperature and the value of the physical parameters associated to time.



This chapter is organized as follows: In Sec. \ref{sec:set-up} we lay out and discuss the assumptions and features of a general cooling protocol. In Sec. \ref{sec:single_inequality} we show recent bounds on cooling protocols with infinite heat baths and catalysts using states out of equilibrium as a resource for cooling \cite{Wilming2017}. In Sec. \ref{sec:masanes} we consider the case of work as a resource for cooling in the presence of a finite heat bath \cite{Masanes2017}. In Sec. \ref{sec:dimension} we present formulations of the UP in terms of the dimension of the heat bath. Lastly, in Sec.\ref{sec:lineal_refrigerators} we consider a less general scenario of more practical relevance by studying the cooling bounds and the UP for networks of harmonic oscillators \cite{Freitas2018}.

\section{General setup for cooling processes}\label{sec:set-up}

In the following we lay out a general framework that includes as particular cases the different types of cooling protocols.
This general cooling process consists of a joint transformation of the following subsystems:
\begin{itemize}

\item \textbf{The system $\S$} is what we want to cool down to the lowest possible temperature.
The system has Hilbert space dimension $d_{\S}$, and its initial and final states are denoted by $\rho_{\S}$ and $\rho'_{\S}$ respectively. The Hamiltonian $H_{\S}$ has ground-space projector $P_\text{gr}$ with degeneracy $g$, and the energy gap above the ground state is $\Delta$.
Most of the following results apply to the case where the system is initially in thermal equilibrium $\rho_{\S} = \omega_\beta (H_{\S})$, at the same temperature than the bath $T = 1/\beta$, where we define the equilibrium state
\begin{align}
  \omega_{\beta}(H) \coloneqq \frac{e^{-\beta H}}{ \Tr (e^{-\beta H})}\ .
\end{align}
The quality of the cooling procedure is quantified by the cooling error
\begin{equation}
  \epsilon =
  1-\tr (\rho'_{\S} P_\text{gr})
  \ ,
\end{equation}
or the final temperature-like quantity
\begin{equation}
  \label{T epsilon}
  T' \geq \frac \Delta
  {\ln(d_{\S} /g \epsilon)}
  \ .
\end{equation}
The unattainability results that are presented in what follows, constitute lower bounds for the quantities $\epsilon$ and $T'$, which prevent them to be zero.

\item \textbf{The bath $\B$} can be seen as the  environment of the system, and as such, it is in thermal equilibrium $\rho_{\B} = \omega_\beta (H_{\B})$ at temperature $T =1/\beta$.
The role of the Bath is to absorb entropy from the system $\S$ contributing to its temperature reduction. The Hilbert-space dimension of the bath can be finite  $d_B$ or infinite. Its Hamiltonian $H_B$ has energy range $J_{\B} = \lambda_\text{max} (H_{\B}) - \lambda_\text{min} (H_{\B})$, where $\lambda_\text{max/min} (H_{\B})$ are its largest/lowest eigenvalue.
The energy range $J_{\B}$ can also be finite or infinite.

\item \textbf{The catalyst $\C$} represents the machine that we use for cooling.
As a tool, its initial and final states must be equal $\rho'_{\C} = \rho_{\C}$, such that, at the end of the protocol it can be re-used in the next repetition of the process (Sec.~\ref{sec:single_inequality}).

\item \textbf{The resource $\R$}
is the fuel that will be consumed in the cooling transformation.
As such, there are no constraints on the final state of the resource $\rho'_{\R}$.
The initial state of the resource must necessarily be not in equilibrium $\rho_{\R} \neq \omega_\beta (H_{\R})$, and its utility increases when increasing its energy and/or decreasing its entropy.
Thermodynamic work can also be seen as a type of resource with conditions on its final state, so that, dumping entropy in $\rho'_{\R}$ is not allowed.



\end{itemize}
Once the subsystems of the cooling protocol have been presented we state now formally some fundamental assumptions that are used in the rest of the chapter.
\begin{itemize}

\item \label{ass:uncorrelation} \textbf{Independence Assumption.} All subsystems are initially in a product state $\rho_{\S} \otimes \rho_{\B} \otimes \rho_{\C} \otimes \rho_{\R}$, and the total Hamiltonian is initially non-interacting $H = H_{\S} + H_{\B} + H_{\C} + H_{\R}$.

\item \label{ass:unitary}\textbf{Unitarity Assumption.} The joint transformation of all subsystems is unitary:
\begin{align}\label{eq:general_transition}
\rho'_{\S} = \Tr_{\B \C \R}\! \left[ U \left(\rho_{\S} \otimes \rho_{\B} \otimes \rho_{\C} \otimes \rho_{\R}\right) U^{\dagger}\right]
\end{align}
where $U$ is a unitary operator.

\item \label{ass:energy_conservation}
\textbf{Energy Conservation}
is the requirement that the global unitary commutes with the total Hamiltonian $[U,H]=0$.
This assumption is considered in Sec.~\ref{sec:single_inequality}.
On the other hand, in Secs.~\ref{sec:masanes} and \ref{sec:dimension}, the unitary operator $U$ is unrestricted.
This energetic imbalance is compensated by an expenditure or generation of work.
In general, this work fluctuates, taking different values in different repetitions of the procedure, or adopting coherent super-positions.
It is important to mention that any non-energy-conserving unitary $U$ can be simulated by an energy-conserving one $V$ acting on a larger compound
\begin{equation}
  U \rho_{\S} U^\dagger = \tr_{\R}\! \left(
  V \rho_{\S} \otimes \rho_{\R} V^\dagger \right)
  \ ,
\end{equation}
where $[V, H_{\S} + H_{\R}] =0$.
For this to be possible, the Hamiltonian $H_{\R}$ and the state of the extra system $\rho_{\R}$ have to be of a particular form \cite{Aberg2013}.

\end{itemize}
Although giving up on the Independence Assumption 
might be of interest, it is ubiquitously assumed in the derivation of bounds and laws of thermodynamics and necessary to obtain usual derivations of the second law of thermodynamics \cite{Jennings2010}.
However, it is important to mention that recent efforts \cite{Bera2017_correlations} are going beyond this framework.
Regarding the Unitarity Assumption 
it is mainly motivated by the formalism of quantum mechanics, which prescribes a unitary evolution for systems evolving under time-dependent Hamiltonians~\cite{Horodecki2013, ResourceTheory}. 

The following table includes the classification of all the unattainability results explained in this chapter (first column).
The ``limiting factor" (second column) contains the physical parameters that need to become infinite in order to achieve absolute zero. These can be: the Hilbert-space dimension of the bath $d_{\B}$, the energy range of its Hamiltonian $J_{\B}$, the heat capacity of the bath $C_{\B}(E)$ (defined in Sec.~\ref{sec:masanes}).
The smaller the value of these parameters is, the further from absolute zero the final state of the system becomes.
The third column tells us which results assume energy conservation (``yes"), and which ones require fluctuating work to compensate for the energetic imbalances (``no").
The fourth column specifies which results assume that the heat bath has finite volume, and which do not.
The fifth column informs us about the thermodynamical resource that fuels the transformation. This can be work, non-equilibrium resources $\rho_{\R}$, or both.
The sixth column tells us which setups include a catalyst and which do not.
\begin{center}
\begin{tabular}{ |c||c|c|c|c|c| }
 \hline
 & limiting factor & $[U,H]=0$ & finite bath & resource & catalyst
 \\ \hline \hline
 Allahverdyan (2011) \cite{Allahverdyan2011} & finite $d_{\B}$ and $J_{\B}$ & no & yes & work & no
 \\ \hline
 Reeb (2014) \cite{MunichLandauer} & finite $J_{\B}$ & no & yes & work & no
 \\ \hline
 Scharlau (2016) \cite{Scharlau2016} & finite $d_{\B}$ and $J_{\B}$ & yes & yes & work & no
 \\ \hline
 Masanes (2017) \cite{Masanes2017} & finite $C_{\B}(E)$ and $W_\text{wc}$ & no & yes & work & no
 \\ \hline
 Wilming (2017) \cite{Wilming2017} &finite resources & yes & no & non-eq. & yes
 \\ \hline
 M\"uller (2017) \cite{Mueller_correlations} & finite catalyst & yes & no & both & yes
 \\ \hline
\end{tabular}
\end{center}

\section{Cooling with finite resources
}\label{sec:single_inequality}

In this section we will summarize the results of Ref.~\cite{Wilming2017}. There, cooling processes are considered which involve arbitrary heat bath and catalyst. The only limiting factor is the size of the resource $\R$, which is assumed to be finite dimensional, and, as we will see, the lowest possible temperature can be compactly expressed as a function of the initial state of the resource $\rho_{\R}$.


We will use the set-up of \emph{catalytic thermal operations} \cite{Janzing00,Horodecki2013,Brandao2015} applied to the task of cooling. For this, consider a thermal bath $\B$ described by state and Hamiltonian $(\omega_{\beta}(H_{\B}), H_{\B})$,   a catalyst $(\sigma_{\C}, H_{\C})$ and a finite-dimensional resource $\R$ described by $(\rho_{\R}, H_{\R})$. We do not impose any restriction on the size or dimensionality of $\B$ and the dimension of $\C$ and allow for arbitrary $H_{\B}, \sigma_{\C}$ and $H_{\C}$. These three systems are brought to interact with a system $\S$ which one aims at cooling and is initially at thermal equilibrium with the thermal bath, that is $\rho_{\S}= \omega_{\beta}(H_{\S})$. By imposing the three assumptions laid out in Sec.~\ref{sec:set-up} ---namely, Independence, Unitarity and Energy Conservation---  and that the catalyst is returned in the same state we obtain transitions of the form
\begin{align}\label{eq:transition_with_catalyst}
\rho'_{\S}\otimes \sigma_{\C} = \tr_{\R\B}\! \left[ U \rho_{\R} \otimes \rho_{\S} \otimes \sigma_{\C} \otimes \omega_{\beta}(H_{\B})U^{\dagger}\right] .
\end{align}
where $U$ commutes with the total Hamiltonian. Note that we demand that the catalyst is returned in the same state and uncorrelated with the system $\S$ that one aims at cooling, in this way, it can be re-used for arbitrary future transitions.
The allowed transitions of the form form \eqref{eq:transition_with_catalyst} have been characterized in Ref.~\cite{Brandao2015} for diagonal states, that is, with $[\rho_{\R}, H_{\R}]=0$ and $[\rho'_{\S}, H_{\S}]=0$. It is shown that a transition is possible if and only if
\begin{align}
\label{eq:second_laws}
S_\alpha(\rho_{\R} \| \omega_\beta(H_{\R})) \geq S_\alpha(\rho'_{\S} \| \omega_\beta(H_{\S}))\quad \forall \alpha \geq 0,
\end{align}
where $S_\alpha$ are so-called \emph{Renyi-divergences}. Note that it is in principle necessary to check an infinite number of conditions ---one for each real value of $\alpha$--- to certify that the cooling protocol is possible. In Ref.~\cite{Wilming2017} it is shown that in the limit of very small final temperature $T'$ the infinite set of conditions reduces essentially to the evaluation of a single function, referred to as \emph{vacancy}, and defined by
\begin{equation}\label{eq:def:vacancy}
\mc{V}_{\beta} (\rho,H):=S(\omega_{\beta}(H)\| \rho),
\end{equation}
where $S$ is the quantum relative entropy defined as $S(\rho\| \sigma) =\tr(\rho\log \rho) - \tr(\rho \log \sigma)$. The vacancy becomes a key quantity in relation with the UP, since it is shown that sufficient and necessary conditions for cooling to sufficiently low $T'$ are given respectively by
\begin{align}\label{eq:thirdlawgeneralintro}
\mc{V}_{\beta}(\rho_R,H_R) - K(\rho_R,H_R,\rho_S,H_S,\beta) &\geq \mc{V}_{\beta}(\rho_S,H_S),\\
\mc{V}_{\beta}(\rho_R,H_R) &\geq \mc{V}_{\beta}(\rho_S,H_S)\label{eq:intronecessary},
\end{align}
where $K (\rho_R,H_R,\rho_S,H_S,\beta) \rightarrow 0$ as $T' \rightarrow 0$.
(See \cite{Wilming2017} for a definition of $K$.)
Hence, in the limit of very cold final states where the UP applies, both inequalities converge to a single one ruled by the vacancy. These conditions can be also re-expressed for the multi-copy case $\rho_R=\rho^{\otimes n}$, where each copy has a local Hamiltonian $h$, to obtain lower bound for the final temperature $T'\geq k (n\mc{V}_{\beta}(\rho,h))^{-1}$ where $k$ is a constant.

It is illustrative to compare these bounds with the actual cooling rates achieved by protocols of Algorithmic Cooling \cite{Schulman1999, Boykin2002, Schulman2005, Raeisi2015}. For example, in the seminal work of Ref.~\cite{Schulman1999} it is considered a cooling protocol like the ones described in Sec.~\ref{sec:set-up} but without heat bath nor catalyst, and simply a resource $\R$ of the form $\rho^{\otimes N}$ with trivial Hamiltonian $h=0$.
This protocol provides a cooling error that decreases exponentially with the size of the resource $\epsilon \propto \exp{(-kn)}$ with $k$ being a constant. In turn, the inequality \eqref{eq:intronecessary} implies a bound of the form $\epsilon \geq C \exp(-Rn)$ with $C$ and $R$ being constants which depend of $\mc{V}_{\beta}(\rho,h)$. This has as an implication that for the case of \emph{i.i.d.}~resources the simple protocol of algorithmic cooling from Ref.~\cite{Schulman1999} offers an exponential scaling which is (up to factors in the exponent) optimal, even within the much larger family of of protocols which employ an arbitrarily large bath and catalyst as considered here.


Lastly, let us briefly mention on the significance that the vacancy rules low temperature cooling. For this, it is illustrative to compare to formulations of the second law which bound the extractable work from a given resource $\rho_R$. As it is well-known, the extractable work $W$ satisfies
\begin{align}\label{eq:optimal_work}
W \leq  F_{\beta}(\rho_R,H_R) - F_{\beta}(\omega(H_R),H_R) \propto S(\rho_R\| \omega_{\beta}(H_R))
\end{align}
where $F_{\beta}(\rho,H)=\tr(\rho H) - \beta^{-1} S(\rho)$ is simply the free energy \cite{Aberg2013, Alicki2004, Popescu2013}. Importantly, note the similarities between the vacancy \eqref{eq:def:vacancy} and the r.h.s.~of \eqref{eq:optimal_work}. This gives a common interpretation for the second and third laws in terms of a common function, the relative entropy, which can be regarded as a distance, or a measure of distinguishability between quantum states. The ordering of the arguments is related with strategies of hypothesis testing to discriminate between both states \cite{Tomamichel2016}. In this way one arrives to the general explanation that the \emph{value} of a given resource is determined by its distinguishability from its thermal state, this is measured with the free-energy for the task of work extraction, and measured with the vacancy for the task of cooling.

\subsection{Work as a resource for cooling}\label{sec:work_as_resource}
It is also possible to incorporate in the framework of this section models of work as a particular case of a resource. As already laid out in Sec.~\ref{sec:set-up}, it is possible to simulate the action of a unitary evolution which inputs work ---that is, with $[U,H]\neq 0$--- by considering an external system, $\R$ in this case, which compensates for the energy imbalance. For this one can follow the approach of Ref.~\cite{Horodecki2013} considering as a resource $\R$ a qubit in state $\rho_{\R}=\ketbra{1}{1}$ and with Hamiltonian $H_{\R}=W \ketbra{1}{1}$. One finds that in this case it is possible to cool down to absolute zero, since the transition
\begin{align}
\omega_{\beta} (H_{\S}) \otimes \ketbra{1}{1} \mapsto \ketbra{0}{0} \otimes \ketbra{0}{0}
\end{align}
is possible whenever $W> \log Z_{\beta}$ where $Z_{\beta}$ is the partition function of $\S$ \cite{Horodecki2013}. Although this seems to be in contradiction with the third law of thermodynamics, we note that this procedure only works if the initial resource $\rho_{\R}$ is exactly pure. If instead we consider slightly noisy work $\rho_{\R}= \epsilon \ketbra{0}{0} + (1- \epsilon) \ketbra{1}{1}$, then perfect cooling is impossible for any $\epsilon > 0$, regardless of the value of $W$ (even if it diverges) \cite{Wilming2017}. In this sense, perfect cooling is only possible if we have already as a resource a state which is not full-rank.

\subsection{Cooling by building up correlations}\label{sec:correlations}

In the description of the catalytic thermal operations of Eq.~\eqref{eq:transition_with_catalyst} we impose that the catalyst is returned in the same state and also uncorrelated with the system being cooled $\S$.
Possible alternatives to this scenario have been recently considered \cite{Mueller_correlations, Sparaciari2017, Wilming_correlations}, where the system $\S$ is allowed to build correlations with the catalyst.
In this way, the l.h.s.~of \eqref{eq:transition_with_catalyst} is substituted by a possibly correlated state $\rho'_{\S\C}$ so that $\tr_S (\rho'_{\S\C})=\sigma_{\C}$. These correlations do not prevent one from re-using the catalyst for subsequent cooling protocols. In particular, suppose that we have a series of uncorrelated systems $\S_1,\ldots,\S_N$ that we want to cool.
One can first apply a cooling protocol using $\S_1$ and $\C$, initially uncorrelated, and produce $\rho'_{\S_1\C}$. Afterwards, the catalyst is re-used together with $\S_2$ for another repetition of the cooling protocol, which is possible regardless of the correlations that $\C$ has established with $\S_1$. Building up correlations in this form is advantageous for implementing cooling processes as shown in Ref.~\cite{Mueller_correlations}. There it is shown that, if two diagonal states $[\rho_{\R},H_{\R}]=[\rho'_{\S},H_{\S}]=0$ satisfy $F_{\beta}(\rho_{\R},H_{\S}) \geq F_{\beta}(\rho'_{\S},H_{\S})$ then there always exist a catalyst and a thermal bath so that
\begin{align}\label{eq:transition_correlated_catalyst}
\gamma_{\S\C}= \tr_{\B }(U\rho_{\R} \otimes \omega_{\beta}(H_{\S}) \otimes \sigma_{\C} \otimes \omega_{\beta}(H_{\B})U^{\dagger})
\end{align}
with $U$ commuting with the total Hamiltonian, $\tr_S(\gamma_{\S\C})=\sigma_{\C}$ and $\tr_{\C}(\gamma_{\S\C})$ arbitrarily close to $\rho'_{\S}$.
This can be used to cool at arbitrarily low temperatures while employing finite resources.
For instance, take $\rho_{\R}$ to be a qubit so that $F(\rho_{\R},H_{\R})>F(\ketbra{0}{0},H_{\S})$ where $\ketbra{0}{0}$ is the ground state of $H_{\S}$.
One can always find such a state $\rho_{\R}$ by making it sufficiently energetic.
Then Eq.~\eqref{eq:transition_correlated_catalyst} implies that it is possible to cool as close to zero temperature as desired just with $\rho_{\R}$ as a resource.
It is natural to ask now if this represents a violation of the UP.
It turns out that the dimension of $\C$ and $\B$, at least this is the case in the construction of Ref.~\cite{Mueller_correlations}, diverge as we approximate better the final zero temperature state. Hence, using the Finiteness Assumption introduced in Sec.~\ref{sec:intro}, one would also require diverging time to implement this protocol. On the other hand it is to date unclear what is the particular scaling of the dimension of $\C\B$ in the optimal construction, hence it is open in this scenario which are the actual bounds relating time and temperature.

\section{Cooling with a bath having finite heat capacity}\label{sec:masanes}

The Finiteness Assumption (Sec.~\ref{sec:intro}) imposes that, within a process lasting for a finite time, the volume of the effective heat bath assisting the transformation must be finite.
In physical setups where this volume is not defined, one can impose, alternatively, the finiteness of the heat capacity or free energy.
We stress that the finiteness of these quantities is independent to that of the Hilbert space dimension.
And in particular, a finite region of a typical heat bath (radiation, air, etc) is described by an infinite-dimensional Hilbert space.

The Finiteness Assumption also imposes that, within a finite time, the amount of work injected into the system and bath must remain finite.
In general, this work expenditure fluctuates, taking different values in different repetitions of the procedure, or adopting quantum super-positions.
Since the UP is a bound on the worst-case cooling time (not the average time), the relevant quantity here is the \emph{worst-case work} (not the average work).
Also, the necessity of considering the worst-case work follows from the observation that, if the worst-case work is not constrained then perfect cooling is possible with a heat bath consisting of a single harmonic oscillator (Sec.~V in \cite{Allahverdyan2011}).


The physical setup considered in \cite{Masanes2017} is the following.
The global unitary $U$ characterizing the transformation
\begin{align}\label{cooling proc}
  \rho'_{\S} = \Tr_{\B} \!
  \left( U \rho_{\S} \otimes \rho_{\B} U^{\dagger}\right)
\end{align}
is not required to commute with the total Hamiltonian $H=H_{\S}+H_{\B}$.
This violation of energy conservation must be compensated by an expenditure or generation of work.
That is, energy that is injected into the system and bath without changing their entropy.
There are different ways to define work in this setup, a standard definition being the average value $\bar W = \Tr\!\left[(H-U^\dagger H U) \rho_{\S} \otimes \rho_{\B} \right]$.
However, as mentioned above, we need to consider the \emph{worst-case work}
\begin{align}\label{eq:def wc work}
  W_\text{wc} = \max_{\ket{\phi_1}, \ket{\phi_2}} \! \big\{ E_2 -E_1 : \bra{\phi_2}U\! \ket{\phi_1} \neq 0 \mbox{ and } H\! \ket{\phi_{1,2}} = E_{1,2}\! \ket{\phi_{1,2}}
   \big\}\ .
\end{align}
That is, the largest transition between the energy levels of $H$ generated by $U$.
Note that this expression only makes sense when the initial state $\rho_{\S} \otimes \rho_{\B}$ has full rank, which is our case.
Finally, we remark that this setup does not include a catalyst, and, wether a catalyst would constitute an advantage is an open problem.

\subsection{Results}

For the sake of simplicity, here we consider the case where the Hilbert space of the system has finite dimension $d_{\S}$ and its initial state is thermal at the same temperature than the bath $\rho_{\S} = \omega_\beta (H_{\S})$.
The general case is analyzed in~\cite{Masanes2017}.
Next we see that, in the context of the UP, a central quantity is the density of states of the bath $\Omega(E)$, that is, the number of eigenvalues of $H_{\B}$ within an energy window around $E$.
This allows to write Boltzman's entropy as $\ln \Omega(E)$.

The most general result of this section is the following.
In any cooling process assisted by a bath with density of states $\Omega (E)$, and using worst-case work $W_\text{wc}$, the ``cooling error" $\epsilon$ satisfies, to leading order,
\begin{equation}
\label{bound 1}
  \epsilon \geq
  \frac {\Omega (E_0)\, e^{-E_0/T}}
  {\tr (e^{-H_{\B}/T})}\ ,
\end{equation}
where $E_0$ is the solution of equation
\begin{equation}
  \label{eq:implicit E0}
  \frac {\partial \ln \Omega (E_0) } {\partial E_0}
  = \frac {\ln(2 d_{\S}/3g)} {W_\text{wc}} \ .
\end{equation}
By ``leading order" it is meant that the bound holds for sufficiently large $W_\text{wc}$.
Equation \eqref{eq:implicit E0} always has a unique solution, provided that the micro-canonical heat capacity of the bath
\begin{equation}
  C_{\B} (E) =
  -\left( \frac {\partial \ln \Omega (E) } {\partial E} \right)^2
  \left(\frac {\partial^2 \ln \Omega (E) } {\partial E^2} \right)^{-1}
\end{equation}
is positive and finite for all $E$.
(An example of system with negative heat capacity is a black hole.)

It is important to mention that bound \eqref{bound 1} can be applied to any thermodynamical transformation $\rho_{\S} \to \rho'_{\S}$ that decreases the rank of the state. Where neither the initial nor the final states need to be thermal. In this more general case we define $d_{\S} = \text{rank}(\rho_{\S})$ and $g = \text{rank} (\rho'_{\S})$, and note that nothing in equations \eqref{bound 1} and \eqref{eq:implicit E0} depends on $H_{\S}$. (This is because we are in the regime $W_\text{wc} \gg \|H_{\S}\|_\infty$.)
In particular, setting $H_{\S}=0$, $d_{\S}=2$ and $g=1$, we arrive at the scenario called Landauer's Erasure (see Sec.~\ref{subsec:landauers}).
This shows that the tasks of cooling and erasing information are essentially equivalent.

In order to understand how this result works, let us apply it to a very general family of baths with density of states $\Omega (E,V) = \exp (a V^{1-\nu} E^\nu)$, where $\nu$ is a free parameter in the ranger $0<\nu <1$.
Note that the associated Boltzman entropy is extensive $\ln \Omega (2E,2V) = 2\ln \Omega (E,V)$.
Substituting this in \eqref{bound 1} and \eqref{eq:implicit E0} we obtain the explicit bound
\begin{equation}
  \label{eq:bath family}
  \epsilon \geq \exp\! \left[ -\frac V T
  \left( \frac {a \nu W_\text{wc}} {\ln(2 d_{\S}/3g)} \right)^{\!\frac 1 {1-\nu}} \right]\ .
\end{equation}
As expected, the larger $V$ and $W_\text{wc}$ are, the lower $\epsilon$ can become.
An interesting observation is that, the faster $\Omega (E)$ grows ($\nu$ closer to 1), the weaker is the bound.
Therefore, we can obtain the most general unattainability result by applying result \eqref{bound 1} to the heat bath with fastest growth of its density of states $\Omega (E)$.
To our knowledge, the system with fastest $\Omega(E)$ growth is electro-magnetic radiation (or any massless bosonic field), whose density of states is of the form written above with parameters $\nu=3/4$ and $a = \frac 4 3 15^{-1/4} \sqrt \pi (c \hbar)^{-3/4}$.
Therefore, at this point, we can obtain a universal unattainability result if we use inequality \eqref{T epsilon} and substitute in our bound \eqref{eq:bath family} with the parameters of electro-magnetic radiation, obtaining
\begin{equation}
  \label{eq:T'VW}
  T' \geq
  \frac {15 c^3 \hbar^3} {\pi^2} \,
  \ln\!^4 \! \left(\frac {2 d_{\S}}{3g}\right) \,
  \frac {T\Delta} {V W_\text{wc}^4} \ ,
\end{equation}
in the regime of large $V$ and $W_\text{wc}^4$.

Finally, let us write an UP in terms of time $t$.
From special relativity we have that $V \leq (ct)^3$, and considering, for example, the linear relation $W_\text{wc} \propto t$, we obtain
\begin{equation}
\label{eq:duration}
  T' \geq \text{const}\,
  \frac 1 {t^7} \ .
\end{equation}
Other setups will have different relations between $V$, $W_\text{wc}^4$ and $t$.
But one can always substitute those in \eqref{eq:T'VW} and obtain a suitable UP in terms of time.

\section{Cooling with a finite-dimensional bath}
\label{sec:dimension}

\subsection{Landauer's Erasure}\label{subsec:landauers}

The aim of Landauer's Erasure is to transform any given state $\rho_{\S}$ to a fixed pure state $\proj{0}$, where the Hamiltonian of the system is trivial $H_{\S} = 0$.
For this to be possible, all the entropy from $\rho_{\S}$ has to be transferred to the bath by consuming work.
Here we consider erasure protocols where any unitary acting on $\S\B$ is allowed, without necessarily commuting with the total Hamiltonian. After tracing out the bath we obtain the final state
\begin{align}
\rho'_{\R} = \tr_{\B} ( U \rho_{\S} \otimes \omega_{\beta}(H_{\B}) U^{\dagger})
\ .
\end{align}
Limitations on the purity of the final state $\rho'_{\S}$ have been investigated in Ref.~\cite{MunichLandauer} where it is shown that
\begin{align}\label{eq:bound_landauer}
\lambda_\text{min}(\rho'_{\S}) \geq e^{-\beta J_{\B}} \lambda_\text{min}(\rho_{\S})\ ,
\end{align}
where $\lambda_{\text{min}}(\rho)$ is the smallest eigenvalue of $\rho$ and $J_{\B} = \lambda_{\text{max}}(H_{\B})- \lambda_{\text{min}}(H_{\B})$ is the energy range of $H_{\B}$.
If one assumes a linear scaling of $J_{\B}$ with the size of $\B$, this provides bounds with a similar scaling of those for algorithmic cooling (see discussion in Sec. \ref{sec:single_inequality}). Note also that the fluctuations of external work applied in the process can be at most $J_{\B}$. The bound~\eqref{eq:bound_landauer} exemplifies that for obtaining perfectly pure states $\lambda_\text{min}(\rho'_{\R})=0$ the norm of the bath's Hamiltonian has to diverge, which will also affect the worst-case work, again in the spirit of the results laid out in Sec.~\ref{sec:masanes}.

\subsection{Other results}

The following two results address the problem of cooling a qubit with Hamiltonian $H_{\S} = \Delta \proj{1}$. In Scharlau's result \cite{Scharlau2016} the initial state of the qubit is $\rho_{\S} = \proj{1}$. But despite being pure, it is not trivial to map it to $\proj{0}$, because only energy-conserving unitaries are allowed in that setup.
This implies that the final temperature of the system is bounded by
\begin{equation}
\label{Scharlau}
  T' \geq \frac {T \Delta} {J_{\B} + T\ln d_{\B}}
  \ .
\end{equation}
In Allahverdyan's result \cite{Allahverdyan2011}, the initial state of the qubit is thermal $\rho_{\S} = \omega_\beta (H_{\S})$ at the same temperature than the bath. This state has more entropy than the one considered in \eqref{Scharlau}, and hence cooling requires more effort. On the other hand, arbitrary unitaries are allowed here, giving
\begin{equation}
  T' \geq
  \frac {T \Delta} {J_{\B}}
  \ .
\end{equation}

\section{Cooling with linear quantum refrigerators}\label{sec:lineal_refrigerators}

\noindent
In this section we explain a rather independent development with the goal of
identifying fundamental limits for cooling on a specific class of quantum
refrigerators. The results presented in the previous sections are based on the
assumption that one has access to arbitrary energy conserving operations
over the compound of subsystems given by the system to be cooled, the thermal
bath, catalyst, and work reservoirs. This is important in order to assess the
ultimate limitations for achieving a particular task, which is cooling in this
case. However, the operations that are actually available in practice are much
more restricted. Thus, it is also relevant to study the fundamental
limitations for less general cooling schemes, that however more closely resemble
experimental settings.

We consider the following family of linear and driven quantum refrigerators.
A central system, an arbitrary network of harmonic oscillators, is connected to
different and independent bosonic thermal reservoirs at different temperatures
(see Figure \ref{fig:model}). The central
network is thus open and can also be driven parametrically, by changing in time
the frequency of each oscillator in the network and the interactions between them.
The goal is to drive the system in order to cool a given thermal reservoir,
extracting energy out of it. Many experimental
cooling techniques can be viewed in this way. For example, during laser cooling
of trapped ions, the internal electronic degrees of freedom are driven by a laser
field and act as a `heat pump' that removes energy from the motional degrees of
freedom, dumping it into the electromagnetic field as emitted photons
.

The proposed model has the virtue of being exactly solvable, without invoking
common approximations for the description of open and driven quantum systems.
Therefore, it is possible to obtain and interpret clear mathematical expressions for
key thermodynamic quantities, like work and heat currents. Despite its
simplicity, this general model of thermal machines displays interesting
features. We will see that the fundamental limit for cooling in this kind of machines
is imposed by a pair creation mechanism analogous to the Dynamical Casimir
Effect (DCE). Also, it will be clear that this process cannot be captured by
standard techniques based on master equations valid up to second order in the
coupling between the central system and the thermal reservoirs.

\subsection{The model}

\begin{figure}[ht]
    \centering
    \includegraphics[scale=.4]{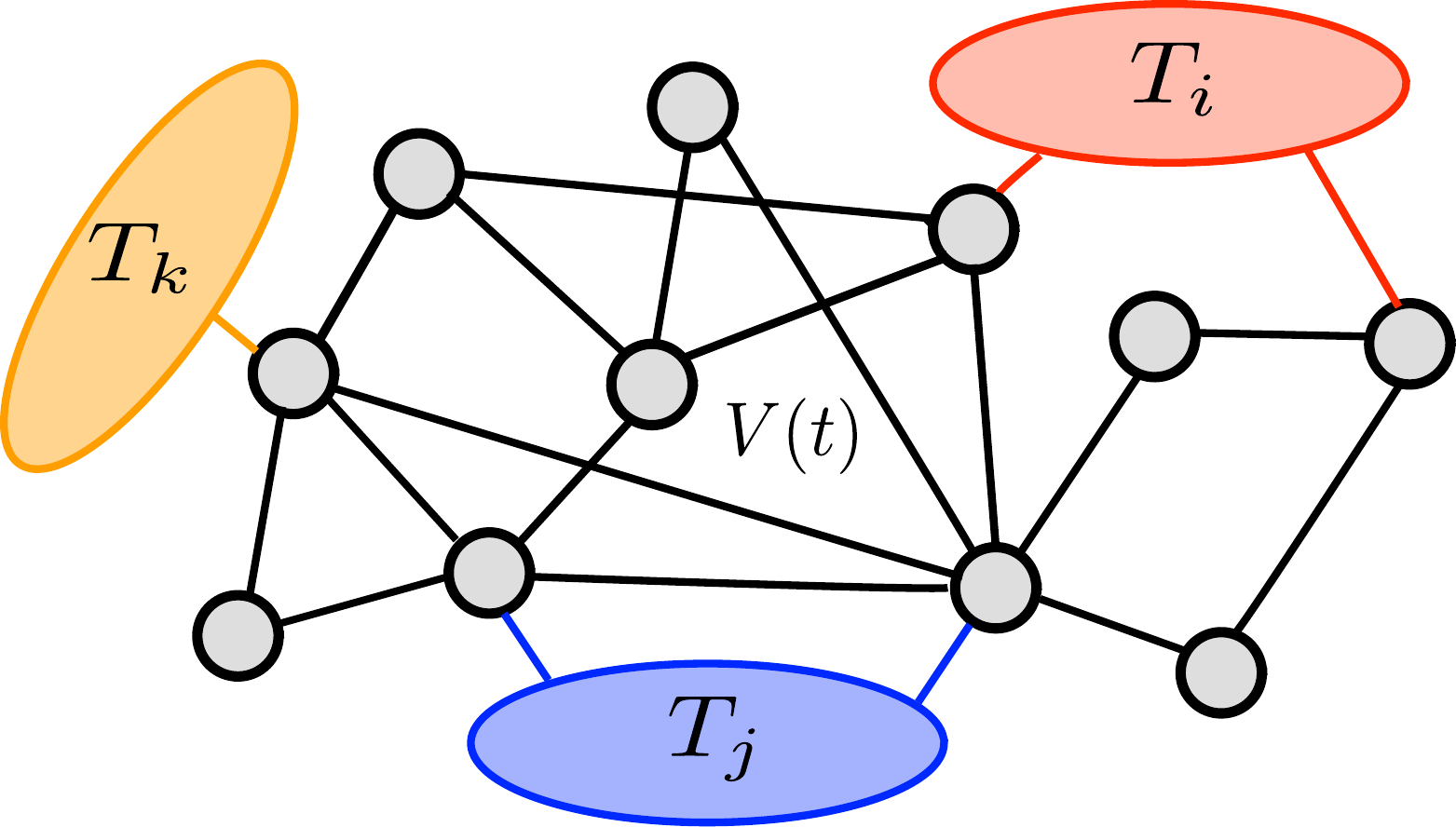}
    \caption{Scheme of the model}
    \label{fig:model}
\end{figure}

\noindent
Figure \ref{fig:model} shows a
scheme of the considered model.
Each black circle represent one of the $N$ quantum harmonic oscillators
composing the network, and links between
them represent bilinear interactions.
The natural frequencies of each
oscillator and the interactions between them can be changed in time. Therefore,
the harmonic network is described by the following quadratic Hamiltonian:
\begin{equation}
    H_\cals(t) = \frac{1}{2} P^T M^{-1} P + \frac{1}{2} X^T V(t) X
    \label{eq:sys_hamiltonian},
\end{equation}
where $X$ and $P$ are vectors whose components are the position and momentum
operators of each oscillator, which satisfy the usual commutation relations, $[X_i,X_j]
=[P_i,P_j]=0$ and $[X_i,P_j]=i\delta_{i,j}$ ($\hbar=1$).
The matrix $M$ has the masses of each oscillator along the diagonal and zeros
elsewhere, while the matrix $V(t)$ encodes the frequencies of each oscillator
and the interactions between them. The variation in time of the matrix $V(t)$
allow us to model an external control that can be performed on the system.

Some nodes of the network are also connected to independent thermal reservoirs.
We will model the reservoirs as collections of harmonic modes which are initially in
a thermal state. Thus, the reservoir or environment $\cale_\alpha$ has a
Hamiltonian
\begin{equation}
    H_{\cale_\alpha} = \sum_{j=1}^{N_\alpha} \frac{\pi_{\alpha,j}^2}{2m}
    + \frac{m\omega_{\alpha,j}^2}{2} q_{\alpha,j}^2
    \label{eq:env_hamiltonian}
    \nonumber
\end{equation}
where the operator $q_{\alpha,j}$ is the position operator of the $j$-th oscillator in the
$\alpha$-th environment, and $\pi_{\alpha,j}$ its associate momentum.
Also, we consider a bilinear interaction between system and reservoirs through the
position coordinates. Thus, for each environment $\cale_\alpha$ we have an
interaction Hamiltonian
\begin{equation}
    H_{\cals,\cale_\alpha} = \sum_{j,k}
    C_{\alpha,jk}\; X_j\:q_{\alpha,k}
    \label{eq:int_hamiltonian},
\end{equation}
where $C_{\alpha,jk}$ are time-independent interaction
constants. Thus, the full Hamiltonian for system and reservoirs is
$H_\calt(t) = H_\cals(t) + \sum_\alpha H_{\cale_\alpha} + \sum_\alpha H_{\cals,\alpha}$.
In the following we will consider cyclic thermodynamic processes for which the driving
performed on the network is periodic. Thus, the function $V(t)$ can be
decomposed in terms of Fourier components $V_k$ as
$V(t)=\sum_k V_k e^{ik\omega_d t}$, where $\omega_d$ is
the angular frequency of the driving.

As we explain below, thermodynamic quantities like heat currents can be obtained
from the state of the central system alone. Thus, if $\rho_T (0) =
\rho_\cals(0) \otimes \rho_\cale(0)$ is an initial product state for the system and the
environment, our main objective is to calculate the subsequent reduced state
for the system:
\begin{equation}
    \rho_\cals(t) = \tr_\cale\left( U(t) \rho_T(0) U^\dagger(t) \right)
    \label{eq:reduced_evolution}
\end{equation}
where the global unitary evolution $U(t)$ corresponds to the Hamiltonian $H_\calt(t)$.
We can do that by solving the equations of motion for the system's operators in the
Heisenberg picture. The linearity of these equations (which
follows from the quadratic structure of the total Hamiltonian) can be exploited to exactly
integrate them in terms of the Green's function of the system. A detailed
explanation of the procedure is given in \cite{freitas2017}. Here, it is enough
to note that since the Hamiltonian is quadratic in the phase space coordinates,
if the full initial state is a Gaussian state, it will remain Gaussian during the
time evolution. Therefore, a complete description of the central system state
$\rho_\cals(t)$ is given by the first
moments $\langle X_i \rangle$ and $\langle P_i \rangle$, and the second moments
$\sigma^{xx}_{i,j} = \langle X_i X_j\rangle$,
$\sigma^{pp}_{i,j} = \langle P_i P_j\rangle$,
and $\sigma^{xp}_{i,j} = \langle X_i P_j + X_j P_i\rangle/2$.
Even if the initial state of the system is not Gaussian, in the regime where the interplay
between the driving and the dissipation induced by the environments determines
a unique asymptotic steady state, this state
will also be Gaussian. Although we will assume in the following that the
system is indeed in such regime, it should be pointed out that in general this will not
be the case, since the driving could give place to parametric resonances,
in which the dynamics is not stable and the memory of the initial state of the
central system is never lost.

As said before, a central object in our treatment is the Green's function
of the harmonic network, which solves its equations of motion and exactly takes
into account the driving and the dissipation induced by the environment.
Explicitly, the Green's function $G(t,t')$ is the $N\times N$ matrix which is the
solution to the following integro-differential equation:
\begin{equation}
    M \frac{\partial^2}{\partial t^2} G(t,t') +
    V_R(t) G(t,t') +
    \int_0^t \gamma(t-\tau) \frac{\partial}{\partial \tau} G(\tau,t') d\tau = 0,
    \label{eq:greens_conf}
\end{equation}
with initial conditions $G(t=t',t') = 0$ and $\frac{\partial}{\partial t} G(t =
t',t') = \mathds{1}_N$. In the previous equation the matrix function $\gamma(t)$,
known as the `damping kernel', takes into account the non-Markovian and
dissipative effects induced by the environment on the network, and
$V_R(t)=V(t)-\gamma(0)$ is a renormalized potential energy matrix.
Specifically, the coefficient $G(t,t')_{j,k}$ encodes the response
of the node $j$ at time $t$, as a result
of a delta-like impulse on node $k$ at time $t'$.
Under the assumptions that the driving $V(t)$ is periodic and that the dynamics
is stable, it can be shown that in the asymptotic regime this function
accepts the following decomposition:
\begin{equation}
G(t,t')=\frac{1}{2\pi}
\sum_k\int_{-\infty}^\infty
d\omega A_k(\omega) e^{i\omega
(t-t')}\, e^{ik\omega_d t},
\label{eq:green}
\end{equation}
where the matrix coefficients $A_k(\omega)$ can be found by solving
a set of linear equations, and can be explicitly calculated in
interesting limits such as the weak driving limit ($|V_k|\ll|V_0|$).
From Eq. (\ref{eq:green}), it is possible to show that for long times the system attains
an asymptotic state which is periodic (with the same period as the driving)
and is independent of the initial state. Also, the second moments
$\sigma_{i,j}^{xx}(t)$, $\sigma_{i,j}^{xp}(t)$, and $\sigma_{i,j}^{pp}(t)$ in
the asymptotic state can be explicitly calculated in terms of $A_k(\omega)$
(see Eq. (\ref{eq:coeff_sigma_xp}) below).

In addition to the Green's function of the network, that characterizes
its dynamics, there are other important quantities that characterize the
reservoirs to which the network is connected. They are the spectral densities
$I_\alpha(\omega)$, one for each reservoir $\cale_\alpha$,
which are $N\times N$ matrices with coefficients defined as
\begin{equation}
 [I_\alpha(\omega)]_{j,k}= \sum_{p=1}^{N_\alpha}
 \frac{1}{m\omega}\;C_{\alpha,jp}\;
 C_{\alpha,kp}\;\delta(\omega-\omega_{\alpha,p}),
 \label{eq:spec_dens}
\end{equation}
where $C_{\alpha,jp}$ are the coupling constants appearing in
Eq. (\ref{eq:int_hamiltonian}).

\subsection{Definition of work and heat currents}

\noindent
We must now define the basic notions of work and heat in our setting.
For this, we can inspect the different contributions to the total time
variation of the energy of the central system, $H_\cals$, which
satisfies
\begin{equation}
\frac{d\mean{H_\cals}}{dt}=
\mean{\partial H_\cals/\partial t}
-i\sum_\alpha\mean{[H_\cals,H_{\cals,\alpha}]}.
\label{eq:energy_der}
\end{equation}
Thus, the variation of the energy
induced by the explicit time dependence
of the system's Hamiltonian
is associated with
work (more precisely, with power), as
\begin{equation}
\dot{\mathcal W}=
\mean{\partial H_\cals/\partial t}.
\end{equation}
In turn,
the variation of the energy of $\cals$ arising
from the interaction with each reservoir
$\cale_\alpha$ is
associated with the heat flowing into
the system per unit time, which we
denote as $\dot
{\mathcal Q}_\alpha$ and turns out
to be
\begin{equation}
\dot{\mathcal Q}_\alpha = -i
\mean{[H_\cals,H_{\cals,\alpha}]}.
\label{eq:def_Q}
\end{equation}
Therefore, equation (\ref{eq:energy_der})
is nothing but the
first law of thermodynamics, i.e.
$d\mean{H_\cals}/dt =
\dot {\mathcal W}+
\sum_\alpha
\dot{\mathcal Q}_\alpha$. In what follows we will
study the average values of the work and
the heat
currents over a driving period (in the
asymptotic regime). These quantities will be
respectively defined as $\dot W$ and
$\dot Q_\alpha$.
Then, the averaged version of the
first law is simply the identity
$0=\dot W+\sum_\alpha \dot Q_\alpha$.
It is interesting to note that an alternative natural
definition for the heat currents could have been given
by the energy change of each reservoir, i.e,
$\dot{\mathcal Q}_\alpha = -\mean{dH_{\cale_\alpha}/dt}$.
As shown in \cite{freitas2017}, in the asymptotic regime
and averaging over a driving period, these two definitions
are equivalent. Thus, the energy lost
by $\cale_\alpha$ is gained by $\cals$ over
a driving period (equivalently, on average, no energy is stored
in the interaction terms).

Introducing the explicit form of the Hamiltonians into Eq. (\ref{eq:def_Q}) it is
possible to arrive at the following expression for the average heat current
corresponding to reservoir $\cale_\alpha$:
\begin{equation}
    \dot{Q}_\alpha = \Tr \left[ P_\alpha \overline{V(t) \sigma^{xp}(t)} M^{-1}\right],
    \label{eq:def_local_heat_cycle}
\end{equation}
where $\overline{X(t)}$ represents the average value of $X(t)$ over a period of
the driving in the asymptotic state, and $P_\alpha$ is a projector over the
sites of the network connected to reservoir $\cale_\alpha$.
In turn, the matrix of position-momentum correlations $\sigma^{xp}(t)$ can be
expressed as
$\sigma^{xp}(t)=\Re\left[\sum_{j,k}S^{xp}_{j,k} \; e^{i\omega_d(j-k)t}\right]$
with:
\begin{equation}
    S^{xp}_{j,k} = \frac{1}{2} \sum_\alpha \int_0^\infty (\omega+k\omega_d)
    A_j(\omega,\omega_d) I_\alpha(\omega) A_k^\dagger(\omega,\omega_d)
    \coth(w/2T_\alpha) d\omega
    \label{eq:coeff_sigma_xp}
\end{equation}
where $T_\alpha$ is the temperature of the initial thermal state of reservoir
$\cale_\alpha$. From these exact results it is possible to derive a physically appealing
expression for $\dot Q_\alpha$, which has a simple and clear interpretation, as
discussed in the following.

\subsection{Heat currents in terms of elementary processes}

It is possible to identify different contributions to the heat current
$\dot Q_\alpha$, and to interpret them in terms of elementary processes
that transport or create excitations in the reservoirs. In \cite{freitas2017}
it is shown that $\dot Q_\alpha$ can be decomposed as the sum of three terms:
\begin{equation}
\dot Q_\alpha=\dot Q_\alpha^\text{RP}+\dot Q_\alpha^\text{RH}+
\dot Q_\alpha^\text{NRH},
\end{equation}
which, respectively, are referred as the `resonant pumping' (RP),
`resonant heating' (RH), and `non-resonant heating' (NRH) contributions.
We will describe below the explicit form of each of these contributions and
their physical interpretation in terms of elementary processes.
The central quantity appearing in the explicit expressions for
$\dot Q_\alpha^\text{RP}$, $\dot Q_\alpha^\text{RH}$ and $\dot Q_\alpha^\text{NRH}$
is the following `transfer' function:
\begin{equation}
    p_{\alpha,\beta}^{(k)}(\omega)=\frac{\pi}{2}
    \Tr \left[
    I_\alpha(|\omega+k\omega_d|) A_k(\omega) I_\beta(\omega)
A^\dagger_k(\omega) \right],
    \label{eq:transfer_func}
\end{equation}
which combines the spectral densities $I_\alpha(\omega)$ (characterizing the
spectral content and couplings of each reservoir) and the coefficients
$A_k(\omega)$ (that determine the Green's function $G(t,t')$ and therefore
characterize the dynamics of the network).
As it will be clear from what follows, the quantity $p^{(k)}_{\alpha,\beta}(\omega)$
can be interpreted as the probability per unit time that a quantum of energy
$\omega$ is removed from $\cale_\beta$ while an quantum of energy
$|\omega+k\omega_d|$ is dumped on $\cale_\alpha$, via absorption (or emission,
depending on the sign of $k$) of an amount of energy equal to $|k\omega_d|$ from
(or to) the driving field.

\begin{figure}
\begin{subfigure}{.5\textwidth}
  \centering
  \includegraphics[height=.55\linewidth]{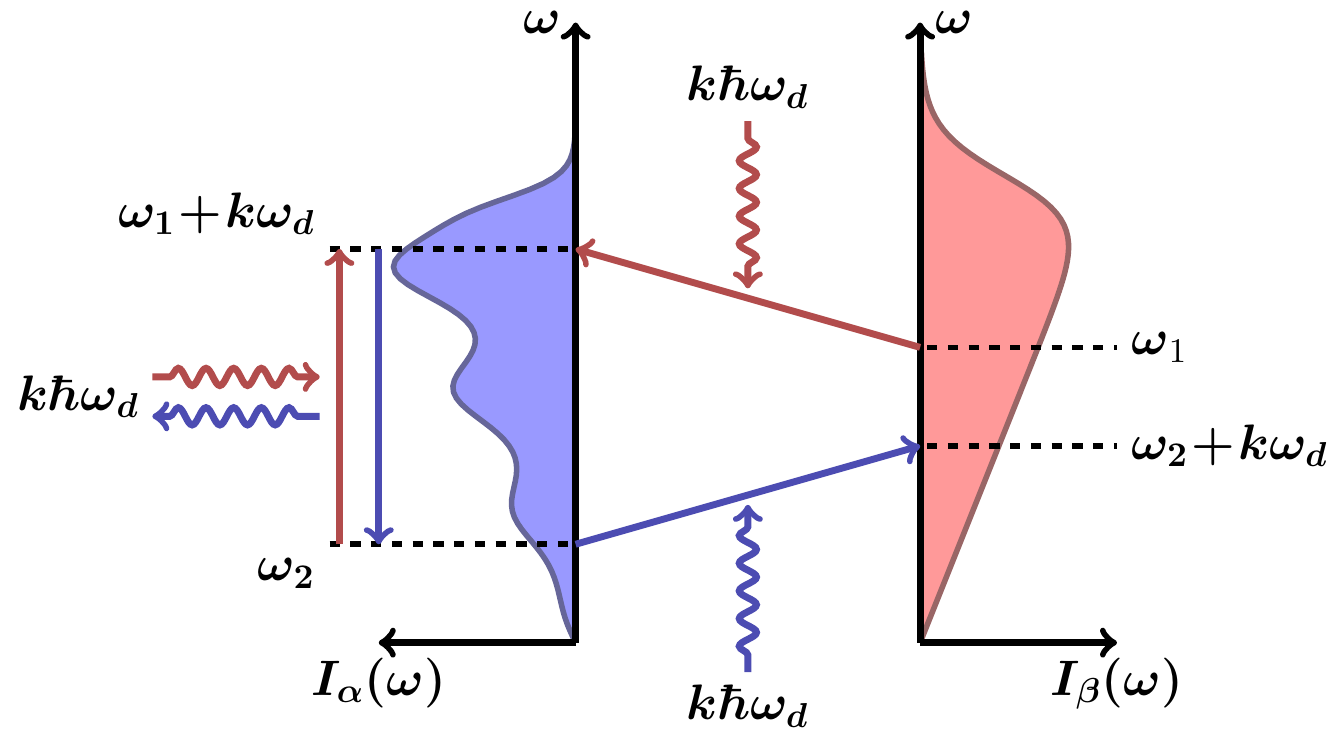}
  \caption{}
  \label{fig:sub1}
\end{subfigure}%
\begin{subfigure}{.5\textwidth}
  \centering
  \includegraphics[height=.55\linewidth]{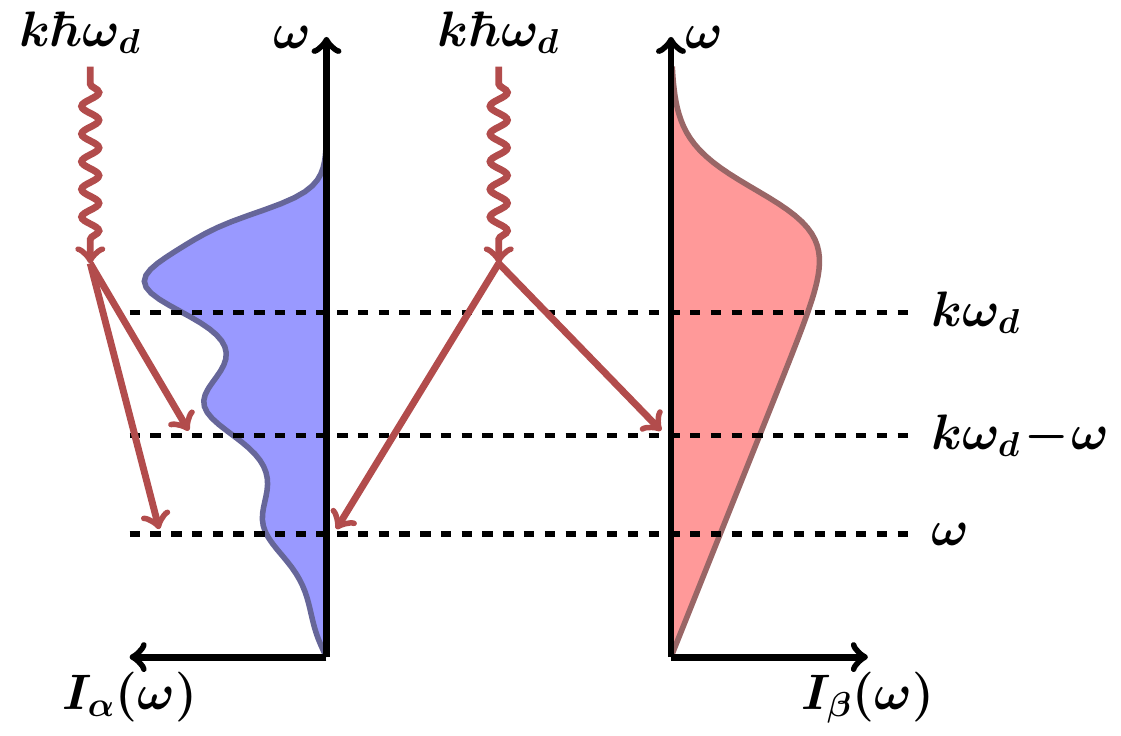}
  \caption{}
  \label{fig:sub2}
\end{subfigure}
\caption{Illustration of the elementary processes contributing to the heat
currents. Two reservoirs $\cale_\alpha$ and $\cale_\beta$ are represented by
their spectral functions. In (a) only the resonant processes are depicted. They
conserve the number of excitations in the environment (but not the energy) and
can involve two environmental modes in different reservoirs (RP) or in
the same reservoir (RH). In the last case, in overall, these processes always
produce heating. In (b) the non-resonant processes are shown:
energy extracted from the driving field is used to simultaneously create
two excitations in the environment. Again, this can involve modes of in the
same or in different reservoirs.}
\label{fig:proc}
\end{figure}

The resonant
pumping (RP) contribution reads:
\begin{equation}
    \dot{Q}_\alpha^{\text{RP}} =
    \sum_{\beta \neq \alpha} \sum_k \int_{0'}^\infty d\omega \; \left[ \omega \;
    p^{(k)}_{\beta,\alpha}(\omega) \; N_\alpha(\omega)
    - (\omega+k \omega_d) \;
    p^{(k)}_{\alpha,\beta}(\omega) \; N_\beta(\omega) \; \right],
    \label{eq:heat_rp}
\end{equation}
where $N_\alpha(\omega) = (e^{\omega/T_\alpha}-1)^{-1}$ is the Planck
distribution at the temperature $T_\alpha$ corresponding to the initial state of reservoir
$\cale_\alpha$ (the Boltzmann constant is $k_b = 1$).
The first term in Eq. (\ref{eq:heat_rp}) is positive and accounts for
energy flowing out of $\cale_\alpha$: a quantum of energy $\omega$ is lost in
$\cale_\alpha$ and excites a mode of frequency $\omega+k\omega_d$ in $\cale_\beta$ after
absorbing energy $k\omega_d$ from the driving. The second term corresponds to
the opposite effect: a quantum of energy $\omega$ is lost from $\cale_\beta$
and dumped into a mode of frequency $\omega+k\omega_d$ in $\cale_\alpha$
after absorbing energy $k\omega_d$ from the driving. These processes are
represented in Figure \ref{fig:proc}-(a). In the same Figure it is shown that
the same processes can take place between two modes of the same reservoir,
which, in overall, always results in heating of that reservoir (since initially
low frequency modes are more populated than high frequency modes and therefore
processes that increase the energy of the reservoir are more probable than
their inversions). Thus, they are considered in the resonant heating (RH)
contribution, which reads
\begin{equation}
    \dot{Q}_\alpha^{\text{RH}} =
    -\sum_k \int_{0'}^\infty d\omega \;
    k \omega_d \;
    p^{(k)}_{\alpha,\alpha}(\omega) \; N_\alpha(\omega)
    \label{eq:heat_rh}
\end{equation}
The lower limit in the frequency integrals of Eqs. (\ref{eq:heat_rp}) and
(\ref{eq:heat_rh}) is $0'= \max\{0, -k\omega_d\}$, since for $k<0$ the mentioned
processes can only take place if the frequency of the arrival mode,
$\omega+k \omega_d$, is greater than zero.

Finally, the last contribution to the heat current is given by the non-resonant
heating term $\dot Q_\text{NRH}$, which for a driving invariant under time
reversal (i.e, such that $V(-t) = V(t+t_0)$), reads:
\begin{equation}
\begin{split}
     \dot{\bar Q}_\alpha^{\text{NRH}} =& - \sum_{k>0} \int_{0}^{k\omega_d}
     d\omega \; k \omega_d \; p_{\alpha,\alpha}^{(-k)}(\omega) \;
     \left(N_\alpha(\omega)+1/2\right) \\
     & -\sum_{\beta \neq \alpha} \sum_{k>0} \int_{0}^{k\omega_d}
     d\omega \; (k\omega_d-\omega)\;p_{\alpha,\beta}^{(-k)}(\omega) \;
     \left(N_\beta(\omega)+1/2\right)\\
     & -\sum_{\beta \neq \alpha} \sum_{k>0} \int_{0}^{k\omega_d}
     d\omega \; \omega \; p_{\beta,\alpha}^{(-k)}(\omega) \;
     \left(N_\alpha(\omega)+1/2\right),
     \label{eq:heat_nrh}
\end{split}
\end{equation}
The physical meaning of this last expression is different than in the previous
contributions. In this case, excitations are not transported among different
modes, but created in pairs from the driving. For example, the first line of Eq.
(\ref{eq:heat_nrh}) takes into account processes in which energy $k\omega_d$
from the driving is used to simultaneously create two excitations in modes
of reservoir $\cale_\alpha$ with frequencies $\omega$ and $k\omega_d - \omega$,
in such a way that their sum equals $k\omega_d$ (note that only terms with $k>0$
enter in the previous expression). The second and third lines of Eq.
(\ref{eq:heat_nrh}) account for processes in which the excitations are created
in modes of different reservoirs, as depicted in Figure \ref{fig:proc}-(b).
Thus, at variance with the RP and RH processes, the ones giving rise to the NRH
contribution do not conserve the number of excitations in the environment.
Consequently, they always produce heating in all reservoirs
(i.e, $\dot Q_\alpha^\text{NRH} \leq 0$).

The only contribution to the heat current capable of describing cooling of
reservoir $\cale_\alpha$ is $\dot Q_\alpha^\text{RP}$. The other two contributions correspond
to processes that end up heating reservoir $\cale_\alpha$ and are always negative.
Thus, to cool this reservoir it is necessary to engineer the driving $V(t)$ or
the spectral densities in order to satisfy the condition
\begin{equation}
\dot Q_\alpha^\text{RP} > |\dot Q_\alpha^\text{RH} + \dot Q_\alpha^\text{NRH}|.
\end{equation}
Let's suppose that all the reservoirs are at the same temperature $T$.
As discussed next, there is always a minimum value of $T$
below which it is impossible to fulfill the previous condition.
Thus, it is impossible to cool reservoir $\cale_\alpha$ below this minimum
temperature.

\subsection{Pairs creation as a limitation for cooling.}
There are other important differences between the resonant and non-resonant
contributions to the heat currents. In first place, we see from Eqs.
(\ref{eq:heat_rp}) and (\ref{eq:heat_rh}) that $\dot Q_\alpha^\text{RP}$ and
$\dot Q_\alpha^\text{RH}$ vanish in the limit of ultra-low temperatures
($T_\alpha \to 0$ $\forall \alpha$). In contrast, $\dot Q_\alpha^\text{NRH}$ does not
vanish but (for $\omega_d>0$) remains constant and negative
in the same limit. This is natural, since in the
ultra-low temperature regime there are no excitations to transport around,
but they can still be created by the driving. Thus, we immediately see
that for sufficiently low temperatures the pair creation mechanism described
above will dominate over the other contributions and will prevent any cooling.

There is an interesting analogy that might help to understand the appearance of
pairs creation in the environment of an open and driven quantum system. In
fact, the integrand in the first line of Eq. (\ref{eq:heat_nrh}) is analogous
to the spectrum of created photons in
the Dynamical Casimir Effect (DCE). The typical explanation of
this effect involves an electromagnetic cavity with periodic boundary conditions.
For example, in a cavity formed by two opposing mirrors, the oscillation of the
mirrors
induces the creation of photon pairs inside the
cavity. In our setting, we can see the driven central system as a periodically
changing boundary condition for the environmental modes. Therefore, it is
natural to expect the creation of excitations pairs in the same way as in the DCE.
The role of the DCE as a fundamental limitation for cooling was, to the best of
our knowledge, first identified in \cite{benenti2015}.

\subsubsection{Pairs creation and the weak coupling approximation}

Another important difference between the contributions $\dot Q_\alpha^\text{RP}$ or
$\dot Q_\alpha^\text{RH}$ on one hand, and $\dot Q_\alpha^\text{NRH}$ on the other hand,
is their scaling with the coupling strength between the central system
and the reservoirs. This can be understood as follows. First,
lets assume that the spectral densities $I_\alpha(\omega)$ are
proportional to some frequency $\gamma$, which typically fixes the rate of the
dissipation that the environment induces on the central system, and is itself
quadratic on the couplings between the system and reservoirs
(see Eq. (\ref{eq:spec_dens})). Also, for simplicity, lets focus in the weak
driving regime ($|V_k| \ll |V_0|$ for $k\neq 0$). In this regime, up to second
order in $V_k$, we have that the matrix coefficients $A_k(\omega)$ in the decomposition
of the Green's function (Eq. (\ref{eq:green})) are given by
$A_k(\omega) = -\hat g(i(\omega+k\omega_d) V_k \hat g(i\omega)$ for $k\neq0$,
where $\hat g(i\omega)$ is the Laplace transform of the Green's function
of the network without driving. Therefore,
the functions $p^{(k)}_{\alpha,\beta}(\omega)$ are:
\begin{equation}
    p_{\alpha,\beta}^{(k)}(\omega)=\frac{\pi}{2}
    \Tr \left[
    I_\alpha(|\omega+k\omega_d|)
    \hat g(i(\omega+k\omega_d) V_k \hat g(i\omega)
    I_\beta(\omega)
    \hat g(-i\omega)  V_k \hat g(-i(\omega+k\omega_d)
     \right].
\end{equation}
These functions are proportional to $\gamma^2$. However, when integrated over
the full frequency range, as in Eqs. (\ref{eq:heat_rp}) and (\ref{eq:heat_rh}),
the result is proportional to $\gamma$. The reason for this is the presence of
poles, or resonance peaks, in the function $\hat g(i\omega)$, whose contribution
depends on the dissipation rate and thus on $\gamma$. Then, the resonant
parts of the heat current, $\dot Q^\text{RP}_\alpha$ and
$\dot Q^\text{RH}_\alpha$, are proportional to $\gamma$.
In contrast, that is not always the case for $\dot Q^\text{NRH}_\alpha$,
since the integration range in the terms of Eq. (\ref{eq:heat_nrh}) is limited
to $k\omega_d$ and might not include any resonance peak of
the functions $p^{(-k)}_{\alpha,\beta}(\omega)$.
As a simple example, if we have a purely
harmonic driving at frequency $\omega_d$ (i.e, we only have Fourier coefficients
$V_{\pm 1}$ and $V_0$, and $|V_{\pm 1}| \ll V_0$), then
$\dot Q_\alpha^\text{NRH} \propto \gamma^2$ for $\omega_d < \Omega_0$, where $\Omega_0$
is the smallest resonant frequency in $p^{(-1)}_{\alpha,\beta}(\omega)$.
Thus, in this situation, the creation of excitation pairs in the environment
is a process of fourth order in the interaction Hamiltonian between system
and reservoirs (recall that $\gamma$ is second order in the interaction
constants). For this reason, it is not captured by master equations that are
derived under the `weak coupling' approximation and are
valid, as is usual, only to second order in the interaction Hamiltonian.

For high temperatures and in the weak coupling regime, the
term $\dot Q_\alpha^\text{NRH}$ can be disregarded in front of
$\dot Q_\alpha^\text{R} = \dot Q_\alpha^\text{RH} + \dot Q_\alpha^\text{RP}$.
However for any fixed value of $\gamma$, no matter how small, there exist
a minimum temperature below which $\dot Q_\alpha^\text{NRH}$ will dominate
over $\dot Q_\alpha^\text{R}$. This minimum temperature will depend on $\gamma$,
and from other details such as the driving protocol and the spectral densities
of the reservoirs. An analysis of the minimum temperature for an adaptive
procedure that was proposed to violate the unattainability
principle\cite{Kurizki} was presented in \cite{freitas2017}.
Also, in \cite{Freitas2018} it is shown that the standard limits
for Doppler and sideband cooling of a single quantum harmonic oscillator
can be derived from this formalism as an special case. This is reviewed in the
next section.

The breakdown of the weak coupling approximation for low temperatures is
known and can also be deduced from the failure of this approximation to capture
quantum correlations between system and environment in that regime
\cite{allahverdyan2012}. However,
our study of this exactly solvable model of driven and open quantum system
allows us to understand what kind of processes are missed by that approximation.
Also, it makes clear that the pair creation process is the one imposing
a minimum achievable temperature for the studied family of driven refrigerators.
As a final comment, we note that the if the pairs creation process
is not taken into account, the validity of the unattainability principle
depends on the properties of the spectral densities\cite{levy2012}.

\subsection{Cooling a single harmonic oscillator}

In this section we employ the formalism explained above to analyze
a simple situation: the cooling of a
single quantum oscillator. Analyzing the
cooling limit for a single oscillator is relevant in several
contexts, such as in the case of cold trapped ions\cite{diedrich1989},
trapped atoms\cite{hamann1998}, or micromechanical oscillators\cite{teufel2011}.
For this we will consider that our working medium $\cals$ is a single
parametrically driven harmonic oscillator
that is in simultaneous contact with two reservoirs.
One of these reservoirs, $\cale_A$ has a single harmonic mode
that we want to cool. The other reservoir, $\cale_B$, is where the energy
is dumped (this reservoir typically represents
the electromagnetic field). As we will see, this model
is an interesting analogy to other more realistic models
for laser cooling. Notably, this simple model is sufficient
to derive the lowest achievable temperatures in the most
relevant physical regimes (and to predict their values in
other, still unexplored, regimes).

Thus, we consider the spectral
density of $\cale_A$ to be such that
\begin{equation}
I_A(\omega)=\tilde I_A\, \delta(\omega-\omega_m).
\end{equation}
where $\omega_m$ is the frequency of the mode to be cooled and
$\tilde I_A$ is a constant
measuring the strength of the
coupling between $\cale_A$ and
$\cals$. In this case,
the frequency integrals needed to obtain the different
contributions to the heat current $\dot Q_A$ are trivial. Clearly, the RH
contribution is absent since $\cale_A$ consists only of a
single mode. The lowest achievable temperature is defined
as the one for which
the heating and cooling terms
balance each other. Using
Eqs. (\ref{eq:heat_rp}) and (\ref{eq:heat_nrh}) it is
simple to compute their ratio as
\begin{equation}
\left| \frac{ \dot{Q}^\text{RP}_A}{\dot Q_A^{NRH}} \right|
   = \frac{\bar n}{1+\bar n}
   \frac{\sum_{k\ge 1}I_B(k\omega_d+\omega_m) |A_k(\omega_m)|^2}{\sum_{k\ge k_d} I_B(k\omega_d-\omega_m)
|A_{-k}(\omega_m)|^2},
    \label{eq:ratiofull}
\end{equation}
where $k_d$ is the smallest integer for which
$k_d\omega_d>\omega_m$ and
$\bar n=N_A(\omega_m)$ is the average number
of excitations in the motional mode.  In order to simplify our analysis,
 we neglected the heating term appearing in the resonant
 pumping current $\dot Q^{RP}$ (i.e, the transport of excitations
 from $\cale_B$ to $\cale_A$). By doing this, we study
 the most favorable condition for
 cooling, assuming that the pumping of
excitations from $\cale_B$ into
$\cale_A$ is negligible. This is equivalent
 to assuming that the temperature of $\cale_B$ is
 $T_B\simeq 0$. Although this is a reasonable approximation
 in many cases (such as the cooling of a single trapped
 ion) we should have in mind that by doing this, the
 limiting temperature we will obtain should be
 viewed as a lower bound to the actual one.
 Thus, the condition defining the lowest bound
 is that the ratio between
the RP and NRH currents is
of order unity. Using the previous expressions,
it is simple to show that this implies that
\begin{equation}
 \frac{\bar n}{\bar n+1}=
 \frac{\sum_{k\ge k_d} I_B(k\omega_d-\omega_m) |A_{-k}(\omega_m)|^2}
{\sum_{k\ge 1}I_B(k\omega_d+\omega_m) |A_k(\omega_m)|^2}.
     \label{eq:nfull}
\end{equation}
To pursue our analysis, we need an expression
for the Floquet coefficients $A_k(\omega)$.
This can be obtained under some simplifying
assumptions. In fact,
if the driving is harmonic
(i.e. if $V(t)=V_0 + V(e^{i\omega_dt}+e^{-i\omega_dt})$)
and its amplitude is small (i.e. if
$V\ll V_0$), we can
use perturbation theory to compute the
Floquet coefficients to leading order in $V$.
In fact,
\begin{equation}
A_{\pm 1}(\omega_m)\approx -
\hat g(i(\omega_m\pm\omega_d))V\,
\hat g(i\omega_m).
\end{equation}
These are the dominant terms
when $\omega_d>\omega_m$ (which
implies that $k_d=1$). For smaller driving
frequencies, which would require
longer equilibration times and involve
longer temporal scales, terms of
higher order in $k$ (which are higher order
in the amplitude $V$) should be taken into
account. Using the above results, we find that
\begin{equation}
 \frac{\bar n}{\bar n+1} =
   \frac{I_B(\omega_d-\omega_m)
|\hat g(i(\omega_d-\omega_m))|^2}
{I_B(\omega_d+\omega_m)|\hat g(i(\omega_d+\omega_m))|^2}.
\label{eq:nmink1}
\end{equation}
It is interesting to realize that
this last expression can be
rewritten as a
detailed balance condition. In fact, this can be
done by noticing that the Planck distribution satisfies
the identity $\bar n/(1+\bar n)=p_{n+1}/p_n$,
where $p_n$ is  the probability for the $n$-phonon
state. Then, equation (\ref{eq:nmink1})
can be rewritten as $P_{heat}=P_{cool}$, i.e.
as the condition for the identity between
the probability of a heating process and the
one of a cooling process.
The cooling probability, $P_{cool}$, is proportional
to the product of
$p_{n+1}$ (the probability of having
$n+1$-phonons in the motional mode),
$|\hat g(i(\omega_d+\omega_m))|^2$ (the probability
for propagating a perturbation with frequency
$\omega_d+\omega_m$ through the work
medium) and $I_B(\omega_d+\omega_m)$
(the density of final states in the
reservoir where the energy of
the propagating excitation
is dumped).
During this process the motional mode
necessarily looses energy. This is the case
because the energy propagating through
$\cals$ is larger than the driving quantum. The extra energy propagating through
$\cals$ is provided by $\cale_A$, that
is therefore cooled.
On the other hand, the heating
probability, $P_{heat}$ is
the product of $p_n$ (the probability for
$n$-phonons in $\cals$),
$|\hat g(i(\omega_d-\omega_m))|^2$
(the probability
for propagating a perturbation with frequency
$\omega_d-\omega_m$ through the work
medium) and $I_B(\omega_d-\omega_m)$
(the density of final states in the
reservoir where the energy is dumped).
In this case, the motional mode necessarily
gains energy because
the energy propagating through $\cals$ is
smaller than $\omega_d$ (the quantum of
energy provided by the driving). The extra
energy is absorbed by
$\cale_A$ , which is therefore heated.
It is interesting to note that this detailed
balance condition is obtained from our
formalism as a simple limiting case.
A more general detailed balance
condition
can be read from Eq. (\ref{eq:nfull}) (which
goes beyond the harmonic, weak driving
or adiabatic approximations).

To continue the analysis it is necessary to give
an expression for $\hat g(s)$ (the propagator of the undriven work medium).
For this we use a semi phenomenological approach
by simply assuming that, in the absence
of driving, the coupling with
the reservoirs induces an exponential
decay of the oscillations of $\cals$.
In this case, we can simply write
$\hat g(i\omega)=1/((\omega-i\gamma)^2-\omega_0^2)$,
where $\gamma$ is the decay rate and $\omega_0$ is
the renormalized frequency of $\cals$.
The same expression is obtained if we assume
that $\cals$ behaves as if it were
coupled with a single ohmic environment
(this is indeed a reasonable
assumption in many cases, which
is equivalent to a Markovian
approximation, but it certainly
requires the back action of
$\cale_A$ on $\cals$ to be negligible in the
long time limit). Inserting this expression
for $\hat g(i\omega)$ into Eq. (\ref{eq:nmink1}),
we can ask what is the optimal
value of the driving frequency $\omega_d$ that minimizes $\bar n$,
for given parameters $\omega_0$, $\omega_m$ and $\gamma$.
As explained in detail in
\cite{Freitas2018}, in this way it
is possible to recover the well known limits
for the regimes of Doppler and sideband cooling. For the case of Doppler cooling,
in which $\gamma \gg \omega_m$, we obtain that the optimal driving
frequency is $\omega_d \simeq \omega_0 - \gamma$ and the corresponding
minimum occupation is:
\begin{equation}
\bar n_\text{doppler}=\frac{\gamma}{2\omega_m}
\frac{\omega_0}{\omega_0-\gamma} \gg 1
\label{eq:limitdoppler}
\end{equation}
under the additional assumption that $\omega_0 \gg \omega_m$ (that is compatible
with optical settings).
In the opposite limit of sideband cooling ($\gamma \ll \omega_m$) we have
that the minimum occupation is achieved for $\omega_d = \omega_0 - \omega_m$
and is:
\begin{equation}
\bar n_\text{sideband} \simeq \frac{\gamma^2}{4\omega_m^2} \ll 1
\label{eq:limitsideband}
\end{equation}
under the same assumptions. However, our treatment is not restricted to these
regimes and can be employed to obtain the optimal driving frequency and minimal
occupation in the general case.

\subsection{The role of pair creation in
laser cooling}

\begin{figure}
\begin{subfigure}{.5\textwidth}
  \centering
  \includegraphics[height=.6\linewidth]{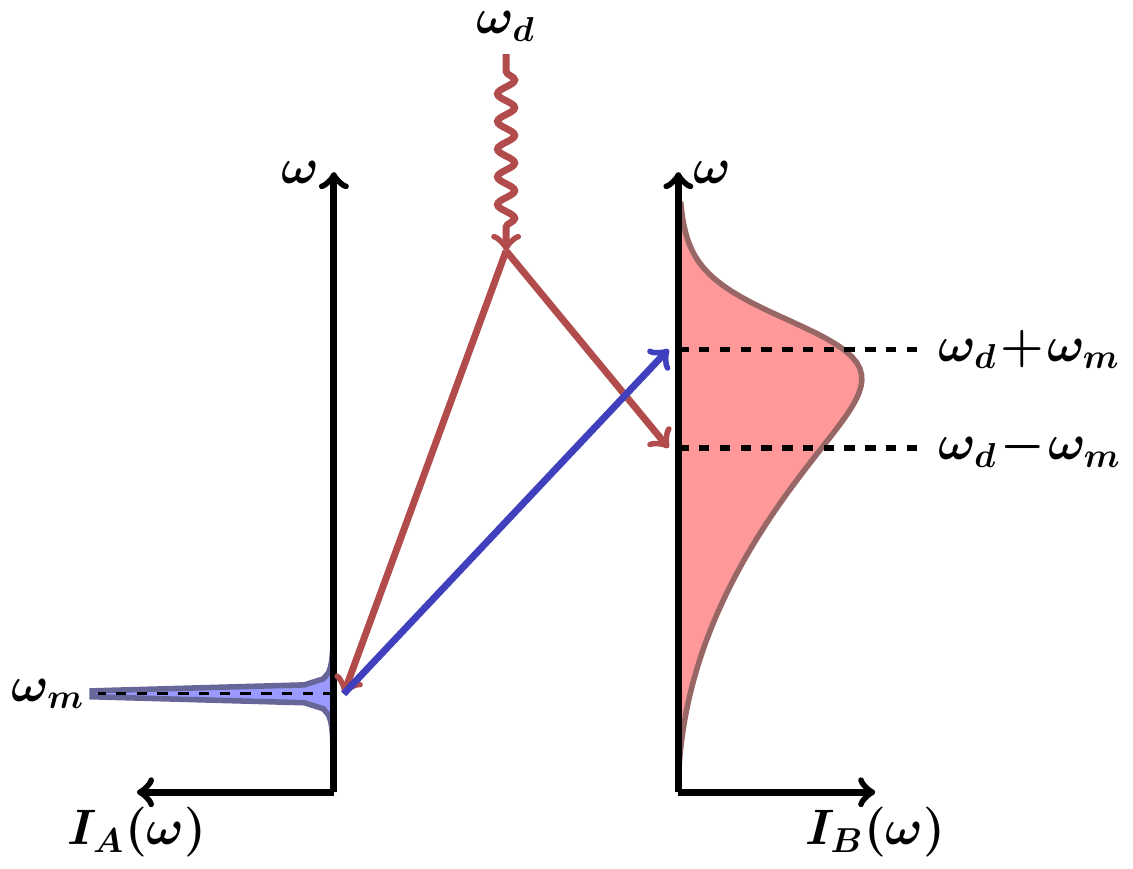}
  \caption{}
  \label{}
\end{subfigure}%
\begin{subfigure}{.5\textwidth}
  \centering
  \includegraphics[height=.4\linewidth]{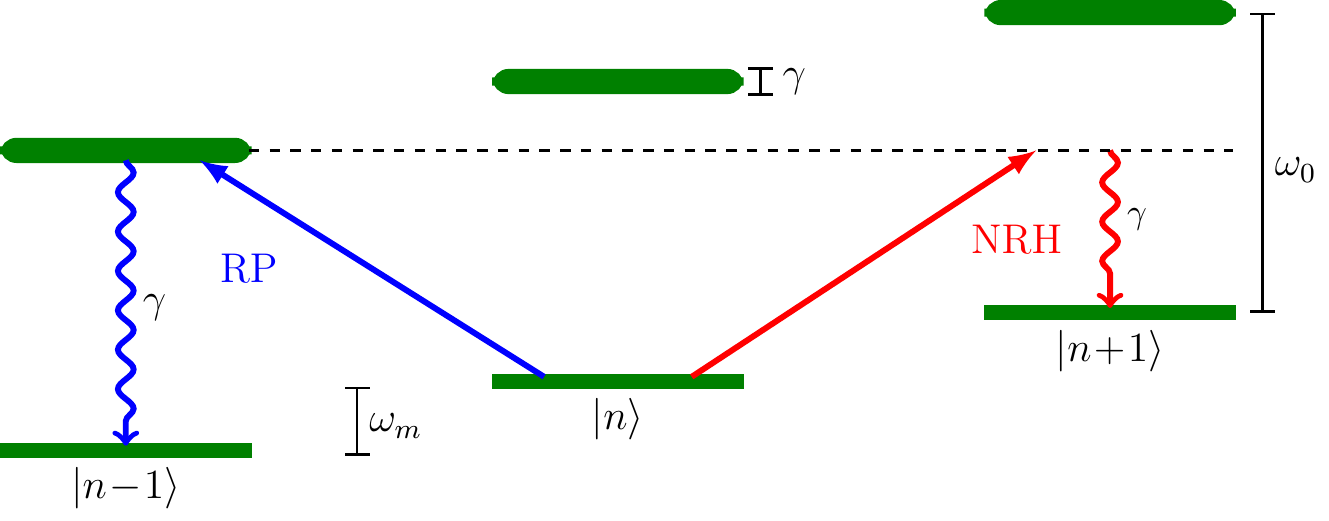}
  \caption{}
  \label{}
\end{subfigure}
\caption{(a) Relevant processes contributing to the heat current of reservoir
$\cale_A$ when $\omega_d>\omega_m$ and $T_B\simeq0$. Energy flows out of
$\cale_A$ and into $\cale_B$
due to a resonant process. Also
energy is dumped into both $\cale_A$
and $\cale_B$ due to non resonant
pair creation.
(b) Usual depiction of the staircase of energy
levels and the transitions between them involved in sideband
resolved laser cooling (actually, there are other non resonant processes in
play, see \cite{eschner2003})
}
\label{fig:laser_cooling}
\end{figure}

According to the previous results,
the origin of the lowest achievable temperature
for the refrigerators we analyzed is imposed
by pair creation from the driving.  This is certainly
not the typical explanation for the reason why
laser cooling stops. However, we will see now
that pair creation has a natural role in laser cooling.
The relevant processes that play a role in the
resonant pumping and non resonant heating
currents are
shown in Fig. \ref{fig:laser_cooling}-(a)
(for $\omega_d>\omega_m$).
Thus, the resonant pumping of energy
out of $\cale_A$ (blue arrow in Figure \ref{fig:laser_cooling}-(a)) corresponds
 to a removal of a motional excitation (a phonon) and its
transfer into the photonic environment. A phonon
with frequency $\omega_m$ disappears
in $\cale_A$ and a photon with frequency $\omega_0$
appears in $\cale_B$. This is possible by absorbing
energy $\omega_d=\omega_0-\omega_m$
from the driving. This process is usually
visualized in a different way in the standard
literature of laser cooling  \cite{eschner2003,marquardt2007,wilson2007},
as shown in Fig. \ref{fig:laser_cooling}-(b).
This Figure shows the energy levels of the
combined system formed by $\cale_A$ and $\cals$.
In our case, both systems are oscillators and
each one of them has an infinite number of
energy levels. However,
we only pay attention to the lowest levels
of $\cals$. Thus, the resonant pumping process
(RP) takes the system from the lowest energy
level of $\cals$ with $n$ phonons into the
excited level of $\cals$ with $(n-1)$ phonons.
Then, as $\cals$ is coupled to the
environment $\cale_B$, it decays from the excited $\ket{e}$
to the ground state $\ket{g}$ by emitting an
excitation (a photon) in $\cale_B$,
whose frequency
is $\omega_0$. This is the key process
responsible for sideband resolved laser
cooling. The system is cooled because resonant
pumping forces the combined $\cale_A-\cals$
system to move down in the staircase of energy
levels.

However, if resonant pumping were the only
relevant process, the above argument would
induce us to conclude that laser cooling could
achieve zero temperature: by going down the
staircase of energy levels, $\cals$ would end
up in its ground state and the motional
mode would end up with $n=0$ phonons.
The reason why this does not happen is the
existence of non resonant
heating. This process is described as NRH in Fig. \ref{fig:laser_cooling}-(a).
It corresponds to the creation of a pair
of excitations consisting of a
phonon and a photon. The phonon has frequency
$\omega_m$ while the photon should have frequency
$\omega_d-\omega_m$. We may choose to
describe this pair creation process as a sequence of
heating transitions that move the combined $\cals$-$\cale_A$
system up along the staircase of energy levels. This can
be done as follows: Suppose that we start
from $n$ phonons in the motional state and
$\cals$ in the ground state $\ket{g}$. Then, $\cals$
can
absorb energy $\omega_d$ from the driving and
jump into a virtual state from which it can decay
back into $\ket{g}$ but with a motional
state with $n+1$ phonons.
This heating transition has the net effect
of creating a phonon and  emitting a photon.
As before,
laser cooling stops (in this sideband resolved limit)
when the resonant cooling transitions are
compensated by non resonant heating transitions
where energy is absorbed from
the driving and is split between two excitations:
one in the motional mode (a phonon) and one
in the environmental mode (a photon). As a
consequence of the non resonant transitions,
the motion heats up. The
limiting temperature is achieved when the
resonant (RP) and non resonant (NRH) processes
balance each other.

Of course, the way in which we are describing
the processes involved in laser cooling (both
the cooling and the heating transitions) is
not the standard one, but provides a new
perspective that allows to draw parallels with
other refrigeration schemes based on external driving.

\bibliography{cooling.bib}

\end{document}